\documentclass[a4paper]{jpconf}  
\usepackage[dvips]{graphicx}
\usepackage{dcolumn}   
\usepackage{bm}        
\usepackage{amssymb}   
\usepackage[utf8]{inputenc}
\usepackage{mathrsfs}
\usepackage{amsmath}
\usepackage[breaklinks=true,colorlinks=true]{hyperref}
\hypersetup{colorlinks=true,citecolor=blue,linkcolor=blue,urlcolor=blue}
\usepackage{textcomp}
\usepackage{subfigure}
\usepackage{color}

\begin{document}

\title{From thin to thick domain walls: An example of\\ the $\varphi^8$ model}

\author{Petr A. Blinov$^1$, Vakhid A. Gani$^{2,3}$ and Aliakbar Moradi Marjaneh$^4$}

\address{$^1$Moscow Institute of Physics and Technology, Dolgoprudny, Moscow Region 141700, Russia}

\address{$^2$Department of Mathematics, National Research Nuclear University MEPhI\\ (Moscow Engineering Physics Institute), Moscow 115409, Russia}

\address{$^3$Theory Department, Institute for Theoretical and Experimental Physics\\ of National Research Centre ``Kurchatov Institute'', Moscow 117218, Russia}

\address{$^4$Department of Physics, Quchan Branch, Islamic Azad university, Quchan, Iran}

\ead{vagani@mephi.ru}

\begin{abstract}
We demonstrate that for some certain values of parameters of the $(1+1)$-dimensional $\varphi^8$ model, the kink solutions can be found from polynomial equations. For some selected values of the parameters we give the explicit formulas for the kinks in all topological sectors of the model. Based on the obtained algebraic equations, we show that in a special limiting case, kinks with power-law asymptotics arise in the model, describing, in particular, thick domain walls. Objects of this kind could be of interest for modern cosmology.
\end{abstract}

\section{Introduction}

Properties of the kink solutions of the $\varphi^8$ field-theoretic model are being actively studied in connection with numerous applications \cite{Lohe.PRD.1979}--\cite{Campos.arXiv.2020}, one of which is the modeling of flat domain walls in cosmology. Moreover, the $\varphi^8$ model is interesting because it admits kinks with both exponential and power-law asymptotics. Moreover, in most cases, the kink solutions of the $\varphi^8$ model can be obtained only in an implicit form, i.e., in the form of the dependence of $x$ on $\varphi$. In recent paper \cite{Gani.PRD.2020}, we have made some progress in obtaining explicit kink solutions for particular values of the model parameters.

In this short paper based on the material presented by Petr Blinov at the ICPPA-2020 conference, we briefly outline the main results that we have recently obtained on the $\varphi^8$ model. We also show how the transition from the case of exponential kink asymptotics to the case of power-law asymptotics \cite{Belendryasova.CNSNS.2019,Manton.JPA.2019,Khare.JPA.2019,dOrnellas.JPC.2020,Christov.PRD.2019,Radomskiy.JPCS.2017,Christov.arXiv.2020,Campos.arXiv.2020} is carried out in the considered version of the $\varphi^8$ model.

\section{The $\varphi^8$ model and its kink solutions}

Consider a field-theoretic model with a real scalar field $\varphi(x,t)$ --- the so-called $\varphi^8$ model --- which can be defined by the Lagrangian
\begin{equation}\label{eq:lagrangian}
	\mathscr{L} = \frac{1}{2} \left( \frac{\partial\varphi}{\partial t} \right)^2 - \frac{1}{2} \left( \frac{\partial\varphi}{\partial x} \right)^2 - V(\varphi)
\end{equation}
with the following polynomial potential:
\begin{equation}\label{eq:potential}
    V(\varphi) = \frac{1}{2}\left(\varphi^2-a^2\right)^2\left(\varphi^2-b^2\right)^2,
\end{equation}
where $a$ and $b$ are constant parameters of the model, $0<a<b$.

The Lagrangian \eqref{eq:lagrangian} is even function of $\varphi$ and has four degenerate minima: $\varphi=\pm a$ and $\varphi=\pm b$. According to this, there are three {\it topological sectors} in the model, $(-b,-a)$, $(-a,a)$ and $(a,b)$, with a kink and an antikink in each sector. A static kink (antikink) is a solution of the first order ordinary differential equation
\begin{equation}\label{eq:eqmo_BPS}
    \frac{d\varphi}{dx} = \pm \sqrt{2V}
\end{equation}
(the plus sign corresponds to a kink, while the minus sign corresponds to an antikink). This equation can be easily solved resulting in the dependence $x(\varphi)$. However, in many situations it is much more convenient to have solution in the explicit form $\varphi(x)$. We managed to obtain explicit solutions for some special cases of the ratio of the parameters $a$ and $b$. For the sake of convenience, below we denote $b/a=n$ and set $b=1$. In this notation, the topological sectors $(-b,-a)$, $(-a,a)$ and $(a,b)$ turn into $(-1,-\frac{1}{n})$, $(-\frac{1}{n},\frac{1}{n})$ and $(\frac{1}{n},1)$, respectively.

\bigskip

\underline{\bf Topological sectors $(a,b)$ and $(-b,-a)$ [or $(\frac{1}{n},1)$ and $(-1,-\frac{1}{n})$]}.

\bigskip

Using the potential \eqref{eq:potential}, from \eqref{eq:eqmo_BPS} we get
\begin{equation}
    x = \frac{1}{2\left(b^2-a^2\right)}\ln\left[\left(\frac{\varphi-a}{\varphi+a}\right)^{1/a}\left(\frac{b+\varphi}{b-\varphi}\right)^{1/b}\right].
\end{equation}
Using the above-mentioned notation $b/a=n$, $b=1$, we obtain
\begin{equation}
\label{eq:main_algebraic_n}
    \left(\frac{n\:\varphi-1}{n\:\varphi+1}\right)^n \frac{1+\varphi}{1-\varphi} = \alpha_n(x),
\end{equation}
where
\begin{equation}
\label{eq:alpha_n}
    \alpha_n(x)=\exp\left[2\left(1-\frac{1}{n^2}\right)x\right].
\end{equation}

\bigskip

\underline{\bf Topological sector $(-a,a)$ [or $(-\frac{1}{n},\frac{1}{n})$]}.

\bigskip

Similarly to the previous case, instead of Eq.~\eqref{eq:main_algebraic_n}, we get the following equation:
\begin{equation}
\label{eq:main_algebraic_aa_n}
    \left(\frac{1+n\:\varphi}{1-n\:\varphi}\right)^n \frac{1-\varphi}{1+\varphi} = \alpha_n(x).
\end{equation}

We have solved algebraic equations \eqref{eq:main_algebraic_n} and \eqref{eq:main_algebraic_aa_n} for $n=2$ and $n=3$, and found explicit formulas for the kink solutions in all topological sectors for these values of $n$.

\subsection{The case $n=2$}

\bigskip

\underline{\bf Topological sectors $\left(-1, -\frac{1}{2}\right)$, $(-\frac{1}{2},\frac{1}{2})$ and $\left(\frac{1}{2}, 1\right)$}:
\begin{equation}\label{eq:kinks_2}
    \varphi_{\rm K}^{(2)}(x) = \cos\left(\frac{1}{3}\arccos\left[\tanh\left(\frac{3}{4}\:x\right)\right]+\frac{\pi m}{3}\right),
\end{equation}
where $m=0,1,2,3,4,5$. At $m=0$ and $m=5$ this equation yields kink and antikink in the topological sector $(\frac{1}{2},1)$; at $m=1$ and $m=4$ --- in $(-\frac{1}{2},\frac{1}{2})$; at $m=2$ and $m=3$ --- in $(-1,-\frac{1}{2})$.

\subsection{The case $n=3$}

\bigskip

\underline{\bf Topological sectors $\left(\frac{1}{3}, 1\right)$ and $\left(-1, -\frac{1}{3}\right)$}:
{
\small
\begin{equation}\label{eq:kinks_3_ab}
    \varphi_{\rm K}^{(3)}(x) = \begin{cases}
        \displaystyle\frac{1}{3} \left(-\sqrt{1-\text{sech}^\frac{2}{3}\left(\frac{8}{9}\:x\right)} \pm \sqrt{2+\text{sech}^{\frac{2}{3}}\left(\frac{8}{9}\:x\right)-\frac{2\tanh \left(\frac{8}{9}\:x\right)}{\sqrt{1-\text{sech}^\frac{2}{3}\left(\frac{8}{9}\:x\right)}}}\right), \quad x<0,\\
        \displaystyle\frac{1}{3} \left(\sqrt{1-\text{sech}^\frac{2}{3}\left(\frac{8}{9}\:x\right)} \pm \sqrt{2+\text{sech}^{\frac{2}{3}}\left(\frac{8}{9}\:x\right)+\frac{2\tanh \left(\frac{8}{9}\:x\right)}{\sqrt{1-\text{sech}^\frac{2}{3}\left(\frac{8}{9}\:x\right)}}}\right), \quad x>0.
    \end{cases}
\end{equation}
}
The plus sign in both formulas corresponds to the sector $\left(\frac{1}{3}, 1\right)$, while the minus sign corresponds to $\left(-1, -\frac{1}{3}\right)$.

\bigskip

\underline{\bf Topological sector $\left(-\frac{1}{3}, \frac{1}{3}\right)$}:
{\small
\begin{equation}\label{eq:kink_3_aa}
    \varphi_{\rm K}^{(3)}(x) = \begin{cases}
       \displaystyle\frac{1}{3} \left(\sqrt{1+\text{csch}^\frac{2}{3}\left(\frac{8}{9}\:x\right)}-\sqrt{2-\text{csch}^{\frac{2}{3}}\left(\frac{8}{9}\:x\right)-\frac{2\coth \left(\frac{8}{9}\:x\right)}{\sqrt{1+\text{csch}^\frac{2}{3}\left(\frac{8}{9}\:x\right)}}}\right), \quad x<0,\\       \displaystyle\frac{1}{3} \left(-\sqrt{1+\text{csch}^\frac{2}{3}\left(\frac{8}{9}\:x\right)}+\sqrt{2-\text{csch}^{\frac{2}{3}}\left(\frac{8}{9}\:x\right)+\frac{2\coth \left(\frac{8}{9}\:x\right)}{\sqrt{1+\text{csch}^\frac{2}{3}\left(\frac{8}{9}\:x\right)}}}\right), \quad x>0,
    \end{cases}
\end{equation}
}

It can be easily shown that the asymptotics of the above kinks are exponential. At the same time, it is interesting to investigate the limit $n\to\infty$, i.e.\ $a\to 0$. In this limit, the potential \eqref{eq:potential} becomes $V(\varphi) = \frac{1}{2}\varphi^4\left(\varphi^2-1\right)^2$, which means the power-law asymptotics of kinks, see, e.g., \cite{Christov.PRD.2019,Radomskiy.JPCS.2017} for details. Let's see how we can come to the power-law kink asymptotics in the limit $n\to\infty$.

\section{Asymptotical behavior and limiting case}

From Eqs.~\eqref{eq:main_algebraic_n} and \eqref{eq:main_algebraic_aa_n} we can get the asymptotics of kinks in the topological sectors $(\frac{1}{n},1)$ and $(-\frac{1}{n},\frac{1}{n})$.

\bigskip

\underline{\bf Topological sector $(\frac{1}{n},1)$}:
\begin{equation}\label{eq:asymptotics_1}
\varphi_{\rm K}^{(n)}(x) \approx \begin{cases}
    \displaystyle\frac{1}{n} + \frac{2}{n}\left(\frac{n-1}{n+1}\right)^{\frac{1}{n}}\exp{\left[\frac{2}{n}\left(1-\frac{1}{n^{2}}\right)x\right]} \quad \mbox{at} \quad x \to -\infty,\\
    \displaystyle 1 - 2\left(\frac{n-1}{n+1}\right)^{n}\exp{\left[-2\left(1-\frac{1}{n^{2}}\right)x\right]} \quad \mbox{at} \quad x \to +\infty.
\end{cases}
\end{equation}

\underline{\bf Topological sector $(-\frac{1}{n},\frac{1}{n})$}:
\begin{equation}\label{eq:asymptotics_2}
\varphi_{\rm K}^{(n)}(x) \approx \begin{cases}
    -\displaystyle\frac{1}{n} + \frac{2}{n}\left(\frac{n-1}{n+1}\right)^{\frac{1}{n}}\exp{\left[\frac{2}{n}\left(1-\frac{1}{n^{2}}\right)x\right]} \thinspace \mbox{at} \quad x \to -\infty,\\
    \displaystyle\frac{1}{n} - \frac{2}{n}\left(\frac{n-1}{n+1}\right)^{\frac{1}{n}}\exp{\left[-\frac{2}{n}\left(1-\frac{1}{n^{2}}\right)x\right]} \thinspace \mbox{at} \quad x \to +\infty.
\end{cases}
\end{equation}

From Eq.~\eqref{eq:asymptotics_1} we see that in the topological sector $(\frac{1}{n},1)$ in the limit $n\to\infty$ at large negative $x$ the argument of the exponent vanishes, which means the transition from exponential to power-law asymptotic behavior of the kink in the sector $(0,1)$.

On the other hand, in the limit $n\to\infty$ the sector $(-\frac{1}{n},\frac{1}{n})$ disappear.

\section{Conclusion}

We have studied the $(1+1)$-dimensional $\varphi^8$ with the potential \eqref{eq:potential} having four degenerate minima. For two particular values of the ratio $b/a=n$ we obtained the explicit formulas for kink solutions. We have also shown that in order to obtain kink solution for any positive integer $n\ge 2$, it is necessary to solve a polynomial equation of $(n+1)$-th degree.

In addition, we have analyzed the asymptotic behavior of the $\varphi^8$ kinks for arbitrary $n$ and considered the case of $n\to \infty$. We have shown that in this limit only two out of three topological sectors remain, and the kinks in these sectors have one power-law asymptotics and one exponential asymptotics. 

In the case of applying the $\varphi^8$ model to describe a cosmological domain wall, the presence of power-law asymptotics means the appearance of a {\it thick domain wall} in the limit $n\to\infty$.

\section*{Acknowledgements}

The work of V.A.G.\ was supported by the Russian Foundation for Basic Research under Grant No.\ 19-02-00930.

A.M.M.\ thanks the Islamic Azad University, Quchan Branch, Iran (IAUQ) for their financial support under the Grant.

The research was also supported by the MEPhI Academic Excellence Project.

\end{document}